\documentclass[a4paper,fleqn,usenatbib,dvipdfmx]{mnras}

\usepackage{newtxtext,newtxmath}

\usepackage[T1]{fontenc}
\usepackage{ae,aecompl}

\usepackage[dvipdfmx]{graphicx}	
\usepackage{amsmath}	
\usepackage{amssymb}	
\usepackage{caption}
\usepackage[subrefformat=parens]{subcaption}
\newcommand{\gcm}{{\rm g}~{\rm cm}^{-3}}

\newcommand{\XHe}{X_{\rm He}}
\newcommand{\XC}{X_{\rm C}}
\newcommand{\XO}{X_{\rm O}}
\newcommand{\TB}{T_{\rm b}}
\newcommand{\FP}{f_{\rm p}}

\title[Models of rotating white dwarfs]
{Rotating white dwarf models with finite-temperature envelope}
\author[S. Yoshida]
{ Shin'ichirou Yoshida
\thanks{E-mail: yoshida@ea.c.u-tokyo.ac.jp}
\\
$^{1}$ Department of Earth Science and Astronomy, Graduate School of Arts and Sciences, The University of Tokyo, \\Komaba 3-8-1, Meguro-ku, Tokyo 153-8902, Japan}
\date{}
\pubyear{201x}

\begin{document}
\label{firstpage}
\pagerange{\pageref{firstpage}--\pageref{lastpage}}
\maketitle

\begin{abstract}
We present new numerical method to compute structures of differentially rotating white 
dwarfs with thermal 
stratification. Our models have cores composed of
ions and completely degenerate electrons and have isentropic envelopes composed
of ions, photons, partially degenerate electrons and positrons.
The models are intended to mimic very early phases
of remnants of white dwarf binary mergers, some of which may lead to type Ia supernovae.
The effect of hot envelope to increase the mass depends on its chemical composition through
the mean molecular weight of the envelope.
For uniformly rotating models, we see only a small increase in mass even in the presence
of hot envelope. Differential rotation changes it drastically and super-Chandrasekhar
mass model whose mass doubles the Chandrasekhar mass of the degenerate star
for some parameter choices. 
We also compute quasi-equilibrium evolutionary sequences of remnants by fixing either
total angular momentum or entropy in the envelope. Existence of these sequences
depends on various factors such as the remnant mass, the profile of differential rotation,
the entropy and the chemical composition of the envelope.
\end{abstract}




\section{Introduction}
\label{sec: introduction}
Mergers of two white dwarfs in binary systems driven by gravitational wave emission
are almost certain to be detected by the planned space-borne laser interferometers
such as Europian-American eLISA \citep{Evans1987, Nelemans2009}, Chinese TianQin \citep{TianQin2016},
and Japanese DECIGO \citep{DECIGO2006}.
The most interesting possibility after mergers is that a sufficiently massive merger remnant
may lead to type Ia supernovae (SNeIa), whose total mass exceeds the Chandrasekhar 
mass of white dwarfs  \citep{Webbink1984, Iben_Tutukov1984}. On the contrary
there have been also claims against the successes of this channel to SNeIa, producing 
neutron stars instead \citep{Saio_Nomoto1985, Nomoto_Kondo1991}.

For smaller mass binaries where SNeIa do not occur, former studies have presented discussions
and evidences that subdwarf O stars, 
R-Corona Borealis stars, and carbon stars may be formed through their mergers
\citep{Webbink1984, Clayton2007, Izzard2007, Longland2011}.
Another interesting possibility is that a white dwarfs with strong magnetic field of $10^6 - 10^9$G
may be created as a remnant of the mergers \citep{Wickramasinghe_Ferraio2000, Garcia-Berro2012}.
It is also intriguing a bimodal distribution in white dwarf mass revealed by {\it Gaia} Data Release 2 may point
to the possibility that some if not all of the massive white dwarfs are formed through the mergers \citep{Kilic2018}.

Considering the richness of binaries and white dwarfs in the Universe, we expect these merging
events are not at all rare. For instance, 
\citet{Rueda2018} estimates the overall cosmic merger rate of white dwarf binaries 
as $(3.7-6.7)\times 10^5 {\rm Gpc}^{-3}{\rm yr}^{-1}$ 
by using an event rate per mass \citep{Maoz2018}
and an extrapolated number of galaxies \citep{Kalogera2001}.

In common merger of a binary with a low mass degenerate member having a larger
radius it is disrupted by unstable Roche-lobe overflow or tidally disrupted to accrete onto a more massive star.
The massive member becomes a core of the remnant whose temperature does not rise initially and
is nearly degenerated. On top of it the debris of lighter star forms a thick and hot envelope, which is heated
up by direct impact or accretion shock. The remnant just after the merger rotates rapidly with strong degree
of differential rotation. As a result the remnant is highly flattened by the centrifugal force.
After the merger, the remnant object evolves through viscous and thermal relaxation.
There have been precedent works mainly focused on these long time evolution 
processes \citep{Yoon2007,Shen2012,Schwab2012,Schwab2016}.
\cite{Shen2012} characterizes the early phase of remnant evolution in three stages. 
After a few dynamical timescale of a remnant (a few second, assuming a normal white dwarf), 
it settles to a (quasi-)hydrostationary state. Then in $10^4$ to $10^8$s, the remnant gradually adjusts its 
differential rotation to a uniform one through the action of viscosity. The timescale may depend
on the so-called $\alpha$ parameter of turbulent viscosity, thus may be regulated by the magneto-rotational
instability\citep{Balbus-Hawley1991a} and/or viscous boundary layer at the boundary between
the core and the envelope. After $10^3$ to $10^4$yrs, the thermal relaxation proceeds
in which the Kelvin-Helmholtz contraction of the envelope takes place
and substantial heat is transported to the core.\\

In this paper we focus on the modeling of the quasi-hydrostationary state of the
remnants before the viscosity smooth out the initial differential rotation.
As the thermal relaxation of the objects is expected to come later, an object
is modeled with two different thermal layers, i.e., the cold core and the hot envelope.
This phase of the remnants lasts for 
a rather short period, only $10^4$s at most \citep{Shen2012}. 
It is, however, interesting to consider remnants in this phase with short duration
since the dynamical processes in and stability of massive remnants 
may be a clue to a prompt explosion of SNeIa \citep{Kashyap2017}.

\section{Formulation}
\label{sec: formulation}

\subsection{Assumptions on the model}
A star we have in mind is a merger remnant of two white dwarfs. According to numerical hydrodynamics 
simulations of white dwarf mergers \citep{Benz1990,Segretain1997,Loren-Aguilar2005,Loren-Aguilar2009, 
Yoon2007,Dan2011,Dan2012,Dan2014,Pakmor2012,Raskin2012,Tanikawa2015,Sato2015,Sato2016}, a low mass member of the binary having a larger radius is disrupted and accretes onto a more massive primary star.
The accreted matter from the secondary is initially heated up by liberating its gravitational
binding energy and forms an extended envelope around the primary. The temperature of the gas may be as high as $10^9$K \citep{Yoon2007} in the envelope and is partially degenerate. The initial temperature range of the remnant may also be estimated by assuming the 
mechanical energy of the binary at the onset of the Roche-lobe filling of the secondary goes
to the internal energy of the formed envelope. This estimate gives the same order of the 
temperature in the envelope.
On the other hand the core remains cold and degenerate until significant thermal conduction
occurs. The timescale may be $10^3$ yrs \citep{Shen2012}, thus we neglect the heating
of the core. Initially the merger remnant is rapidly rotating with high degree of differential
rotation and the viscosity let the rotational profile evolve to that of uniform rotation
in $10^4$ to $10^8$s. We are interested in these initially differentially rotating remnants
which lasts much shorter timescale than that of the thermal evolution.
The rotational profile has a nearly uniformly rotating core with a differentially rotating
envelope (see e.g., Fig.6 in \cite{Yoon2007} ; Fig.1 in \cite{Schwab2012} ; Fig.1 in \cite{Shen2012}).
As for the envelope, we assume it isentropic for simplicity. The hot envelope with temperature
of $10^9$K radiatively cools down. They tend to develop convection \citep{Loeb_Rasio1994}
which leads to the nearly isentropic (neutrally stable to convection) thermal
structure \citep{Kippenhahn_Weigert_book1994}.
Some of the hydrodynamic simulations of white dwarf mergers
also supports this simplification, in which entropy profile is nearly
flat in the outer region of the remnants except at their surface
(see Fig.1 in \cite{Zhu2013},
and Figs.2 and 7 in \cite{Schwab2012})


\subsection{Equation of state of degenerate core}
As the equation of state (EOS) of completely degenerate core, we use the analytic 
expression of pressure and density
as a function of dimensionless Fermi momentum $x$ \citep{Shapiro-Teukolsky}.
It should be noted that the correction due to ion
	Coulomb interactions are important in realistic white dwarfs
	(e.g. Sec.2.4 of \citet{Shapiro-Teukolsky}). We neglect the correction
	in this paper for simplicity.
Mass density $\rho$ and pressure $p$ are related as
\begin{eqnarray}
 \rho &=& bx^3\\
 p &=& a\left(x(2x^2-3)(x^2+1)^{1/2}
+3\ln\left(x+\sqrt{1+x^2}\right)\right).
\end{eqnarray}
Here $x$ is the dimensionless Fermi momentum of electron, $x=p_{_F}/m_ec$
where $p_{_F}$ is the Fermi momentum, $m_e$ is the mass of electron
and $c$ is the speed of light. The value of $a$ and $b$ are
\begin{eqnarray}
 a &=& 6.00\times 10^{22} \quad {\rm dyn/cm^2}\\
 b &=& 9.82\times 10^5\mu_e \quad\gcm,
\end{eqnarray}
where $\mu_e$ is mean molecular weight of electron. 
Enthalpy $h$ is given by
\begin{equation}
 h = \frac{8a}{b}\sqrt{1+x^2}.
\end{equation}

\subsection{Equation of state of hot envelope matter}
For the EOS of hot envelope matter, we adopt the so-called 'Helmholtz' EOS 
by \cite{Timmes-Swesty2000}.\footnote{We use the Fortran subroutine codes and data tables
provided at {\tt http://cococubed.asu.edu/code\_pages/eos.shtml}.} 
The EOS assumes the gas is composed of several species of ions, photons, electrons
and positrons. The electrons and positrons have arbitrary degrees of degeneracy
and relativity. The numerical EOS is constructed by forcing
the thermodynamical consistency when interpolating Helmholtz's free energy.
Given mass density, temperature, mass fraction of ion species, the original subroutine computes 
various thermodynamical quantities such as pressure, entropy, internal energy density, sound
speed, and adiabatic exponents. Whenever other combination of thermodynamical variables are 
needed as input values, we develop wrapper routines that iteratively solve for the variables
by calling the original subroutine.

\subsection{Basic equations of hydrostationary configurations}
Basic equations of hydrodynamics assuming stationary and axisymmetric equilibrium are
the momentum equation
\begin{equation}
	-\frac{1}{\rho} \nabla p - \nabla\Phi - R\Omega(R)^2 = 0,
	\label{eq: momentum}
\end{equation}
and the Poisson equation for gravitational potential,
\begin{equation}
 \Delta\Phi = 4\pi G\rho.
\label{eq: poisson}
\end{equation}
$\Omega$ is the rotational angular frequency of the star which is a function of cylindrical radius
$R$ defined as $R = r\sin\theta$ where $r$ and $\theta$ are the spherical
polar coordinate whose origin is that of the star.
We neglect the effect of general relativistic gravity in our models, which may significantly
affect the result for very massive stars 
\citep{Boshkayev2011,Boshkayev2014,Mathew_Nandy2017,Carvalho2018}. 
In a domain of a star in which fluid is regarded as barotropic (i.e., density depends only on pressure),
Poincar\`{e}-Wavre theorem \citep{Tassoul1978book} tells us that the rotational angular frequency 
depends only on the distance from the rotational axis. Since our stellar models have a degenerate 
core (zero entropy core) and an isentropic envelope, the angular frequency satisfies the condition of the theorem.
We further assume that the angular frequency distribution is continuous at the core-envelope boundary for simplicity.

In this paper physical quantities are normalized as follows (variables with tilde
are dimensionless).
\begin{eqnarray}
 \frac{\rho}{\rho_{\rm c}} &=& \tilde{\rho}\\
 \frac{p}{p_{\rm c}} &=& \tilde{p}\\
 \frac{\Omega}{\sqrt{4\pi G\rho_{\rm c}}} &=& \tilde{\Omega}\\
 \frac{r}{R_\star} &=& \tilde{r}\\
 \frac{\Phi}{R_\star^2~4\pi G\rho_{\rm c}} &=& \tilde{\Phi}
\end{eqnarray}
Here $R_\star$ is the equatorial surface radius, $\rho_{\rm c}$ is density
at the origin and $p_{\rm c}$ is pressure at the origin. With this
normalization we introduce dimensionless parameter $\beta$
in the theory
\begin{equation}
 \beta = \frac{p_{\rm c}}{4\pi G \rho_{\rm c}^2 R_\star^2}.
\end{equation}
Dimensionless enthalpy is defined as
\begin{equation}
\tilde{h}:=\int \frac{d\tilde{p}}{\tilde{\rho}}.
\end{equation} 
Note that a term in the equation containing dimensional enthalpy should 
be multiplied by $\beta$. In the following we omit tilde from
dimensionless quantities.
The momentum equation Eq.(\ref{eq: momentum}) is integrated to result in
\begin{equation}
 \beta h + \Phi - \int^R_0 u\Omega(u)^2 du = C,  
 \label{eq: bernoulli}
\end{equation}
where $C$ is a constant of the first integral (normalized by
$4\pi G\rho_{\rm c}^2 R_\star^2$) of the momentum equation.
Notice that $\Omega$ is a function of $u=r\sin\theta$.
 Different values of the constant $C$ are assigned to each domain of isentropy. 
 The Poisson's equation is cast into an integral equation by using
an appropriate Green's function $G(\vec{r},\vec{r'})$,
\begin{equation}
	\Phi(\vec{r}) = \int G(\vec{r},\vec{r'}) \rho(\vec{r'}) dV',
	\quad G(\vec{r},\vec{r'}) \equiv -\frac{1}{4\pi |\vec{r'}-\vec{r}|}.
		\label{eq: poisson integrated}
\end{equation}

\subsection{Rotation laws\label{sec: rotation laws}}
For differentially rotating barotropic-star models, different analytic forms of 
rotational profile (i.e.. functional forms of angular frequency) are proposed 
in the literatures (see e.g., \cite{Eriguchi_Mueller1985,HSCF}).

We present results of three representative rotational profiles compatible 
with barotropic fluid.
In all cases we introduce dimensionless parameter $k_0$
by which the last term in the left hand side of Eq.(\ref{eq: bernoulli})
is written as
\begin{equation}
 	\int_0^R\Omega(u)^2 u du \equiv k_0\Psi(R).
 	\label{eq: rotational integral}
\end{equation}
The constants in $\Psi$ for each case is chosen in such a way that $\Psi(R=0)=0$.
   \subsubsection{Uniform rotation}
   In this case we define $k_0  = \Omega_0^2$ as a constant
which amounts to the rotational frequency on the symmetry
axis (and throughout the fluid). $\Psi$ is given as,
\begin{equation}
 \Psi = \frac{1}{2}R^2.
\end{equation}
%

  \subsubsection{Slowly rotating core remnants -- "Yoon07"\label{sec: rotation law Yoon}}
\begin{figure}
\includegraphics[width=9cm]{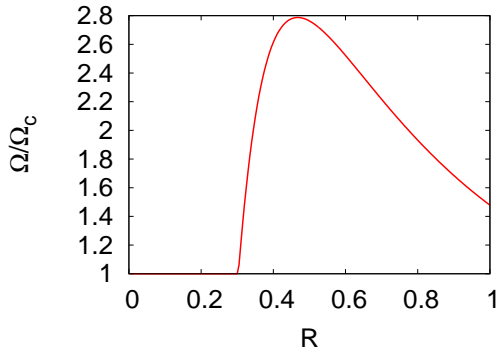}
\caption{A characteristic profile of angular frequency profile of a star with a slowly
rotating core (Eq.(\ref{eq: Omega Yoon07})). The distance from the rotational axis
is normalized by the equatorial radius of the star and the angular frequency is normalized by the that of
the uniformly rotating core. In this normalization $r_{\rm b}=0.3$ and $\epsilon_b=3\times 10^{-2}$ for this
example.
}
\label{fig: Omega Yoon07}
\end{figure}

  \cite{Yoon2007} investigates a secular evolution of a remnant of carbon-oxygen white dwarf merger.
  Their initial model of the remnant consists of slowly and uniformly rotating core and a rapidly 
  and differentially rotating envelope.
  Figure 6 of \cite{Yoon2007} gives the angular frequency profile of the remnant. The profile is
  well-approximated with a simple function,
  \begin{equation}
	\Omega = 
		\begin{cases}
			\sqrt{k_0} & (R\le r_{\rm b})\\
			\sqrt{k_0}\frac{r_{\rm b}^3}{\epsilon_{\rm b}^{5/4}}R^{-3}\left(R-b\right)^{5/4} & (R\ge r_{\rm b}),
		\end{cases}
		\label{eq: Omega Yoon07}
   \end{equation}
where $r_{\rm b}$ is the equatorial core radius, $\epsilon_{\rm b}$ is a small parameter making
the angular velocity $\Omega$ continuous at $r_{\rm b}$,
and $b\equiv r_{\rm b}-\epsilon_{\rm b}$ (Fig.\ref{fig: Omega Yoon07}).  We name the rotation law
"Yoon07". The integral $\Psi$ is written as
	\begin{equation}
		\Psi =
			\begin{cases}
				\frac{k_0}{2}R^2 & (R\le r_{\rm b})\\
				k_0\left(F(R)-F(r_{\rm b}) + \frac{r_{\rm b}^2}{2}\right) & (R\ge r_{\rm b})
			\end{cases}
			\label{eq: Psi Yoon07}
	\end{equation}
where $F(R)  \equiv \frac{5\tan^{-1}\left(\frac{\sqrt{R-b}}{\sqrt{b}}\right)}{64b^{3/2}}
				+ \frac{\sqrt{R-b}(-48b^3+136b^2R-118bR^2+15R^3)}{192bR^4}$.
The uniformly rotating core, formerly being a primary star, is slowly rotating while
the envelope has a large angular momentum inherited from the orbital angular momentum
of the secondary. Outside the core, there is a small region where the angular frequency increases
as a function of $R$. $\Omega$ in the outer envelope monotonically decreases.  
  \subsubsection{Fastly rotating core remnants - "Kepler"\label{sec: rotation law Kepler}}
	A rotation law that resembles that of Keplerian flow ($\Omega \sim R^{-3/2}$) in the envelope may be defined as
	\begin{equation}
		\Omega = \frac{\sqrt{k_0}}{R^{3/2}+D^{3/2}},
	\end{equation}
	that gives $\Psi = \psi(R)-\psi(0)$, where
	\begin{equation}
		\psi = \frac{1}{9}\left(-\frac{6\sqrt{R}}{D^{3/2}+R^{3/2}}
		-\frac{2\sqrt{3}\tan^{-1}\left(\frac{1-\frac{2\sqrt{R}}{\sqrt{D}}}{\sqrt{3}}\right)}{D}
				+ \frac{\ln\frac{(\sqrt{D}+\sqrt{R})^2}{D^2-D R+R^2}}{D}\right)
	\end{equation}
We hereafter call this rotation law "Kepler".
For $R\ll D$, the profile limits to that of the uniform rotation. In the other
extreme, $R\gg D$, $\Omega\sim R^{-3/2}$. In general $D$ is not necessarily the radius
of the core $r_{\rm b}$, but we set $D=r_{\rm b}$, which means the rotational profile change
its slope at $r_{\rm b}$. In this rotational profile the core is rotating more rapidly than the envelope
whose rotational profile asymptotes to the Keplerian.
This profile is motivated by the result of a merger simulation by \citet{Shen2012} (see their Figs.1 and 3)
in which the accreted matter from the secondary forms a hot Keplerian disk around the primary.
Apart from the narrow transition region, the core rotates faster than the Keplerian envelope
in their result. Although the profile is different from that of \citet{Yoon2007}, we explore this possibility as well.

\subsection{Our self-consistent field method
\label{sec: HSCF}}
\cite{HSCF1,HSCF2} presented a versatile method (Hachisu's self-consistent field (SCF) method, HSCF method) of solving 
stellar equilibrium configurations that stably solves 
highly deformed equilibrium configuration of a single star or binary stars. The method gives convergence to rapidly 
and differentially rotating stellar models close to mass-shedding limit for general barotropic equation of state with 
various stiffness. By using HSCF \cite{Hayashi1998} studied differentially rotating hot cores of massive stars ("fizzlers") in which
they took into account finite temperature effect. They fix entropy and electron fraction
so that the EOS is mimicked by a barotropic one. 
\cite{Fujisawa2015} developed an extension of the HSCF to finite temperature and non-barotropic stars by assuming shellular rotation
law in which angular frequency is an analytic function of radial coordinate. The method improved 
a formerly-developed numerical method that directly solves a descretized set of partial differential equations 
by using Newton-Raphson scheme \citep{Uryu1994,Uryu1995}.

Here we present a new SCF formalism to compute rapidly rotating multi-temperature stellar model 
based on the assumptions in the previous sections. 
Our stellar model consists of two zones each of which is barotropic (zero entropy in the core and isentropic
in the envelope). Thus there appear two constants of Eq.(\ref{eq: bernoulli}), which we name $C_{\rm in}$ and $C_{\rm out}$.
$C_{\rm in}$ is for the degenerate core and $C_{\rm out}$ is for the hot envelope. Other parameters to be fixed 
are $\beta$ and $k_0$. 

\begin{figure}
\includegraphics[width=9cm]{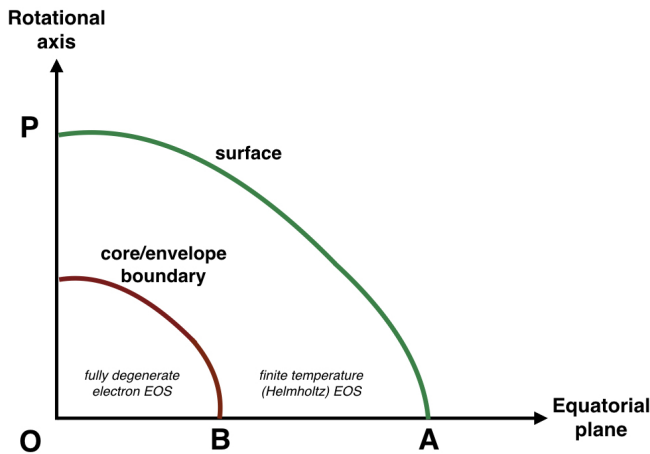}
\caption{Schematic figure of the quadrant of the meridional section of the star. We assume the star
is axysymmetric as well as symmetric with respect to its equatorial plane. Point 'O' is the origin, at which
star's center of mass resides. Point 'A' is at the equatorial surface of the star. Point 'P' is an intersection
of the symmetric axis of the star with the surface. Point 'B' is a point on the core-envelope boundary
on the equatorial plane.}
\label{fig: schematic HSCF}
\end{figure}

The schematic configuration of the star is shown in Fig.\ref{fig: schematic HSCF}.
Point 'O' is the origin, at which
star's center of mass resides. Point 'A' is at the equatorial surface of the star. Point 'P' is an intersection
of the symmetric axis of the star with the surface. 'B' is a point on the core-envelope boundary
on the equatorial plane. Our SCF type iterations are done as follows (procedure [a] to [h]).
 
\begin{enumerate}
\renewcommand{\labelenumi}{[\alph{enumi}]}
\item First the model parameter is provided. They are the central density ($\rho_{\rm c}$), the ratio of the pressure at B
to the central pressure ($\FP$), the envelope temperature at B ($\TB$), the polar and equatorial axis ratio 
($a_X\equiv \overline{OP}/\overline{OA}$), the chemical composition of the core and the envelope. Notice that the point B
is not initially specified, but $\FP$ is. 
\item The function $\Psi$ of rotational integral (Eq.\ref{eq: rotational integral}) is computed on the grid points.
\item Enthalpy distribution on the computational grid points is initially guessed.
\item At each grid point density is computed through the EOS routines. 
\item The density distribution is plugged into Eq.(\ref{eq: poisson integrated}) to
compute the gravitational potential $\Phi$. 
\item The distance $r_{\rm b}$ between O and B, $\beta, k_0, C_{\rm in}, C_{\rm out}$ are iteratively solved. The five equations solved
are, Eq.(\ref{eq: bernoulli}) evaluated at O, A, P, and at B for $r= r_{\rm b}-0$ (core),
as well as the continuity of pressure at B, i.e., $p(r_{\rm b}-0,\pi/2)=p(r_{\rm b}+0,\pi/2)\equiv p_b$. The boundary between the core
and the envelope is specified by the ratio $\FP$. We use Newton-Raphson
method to solve for these parameters. 
\item Once the parameters above are fixed, we compute the enthalpy at each grid points by Eq.(\ref{eq: bernoulli}).
The surface of the star is identified as zeroes of pressure.
\item If the residual of enthalpy from that of the former iteration is small enough, we have a solution. Otherwise we
continue the iterations.
\end{enumerate}
The quality of a converged solution is measured by the virial check \citep{HSCF1}, $VC$,
\begin{equation}
	VC \equiv \left|2\int\rho (R\Omega)^2 dV + \int \rho\Phi dV + 6\beta\int p dV\right|.
\end{equation}
For grid numbers of $(r,\theta)=(200,50)$ we have $VC={\cal O}[10^{-3}]$, compared to 
${\cal O}[10^{-4}]$ for the original HSCF. The slightly degraded quality of the current method
reflect the discontinuity of such thermodynamic variables as $\rho$ at the core-envelope
boundary. Though pressure is made continuous there, density is discontinuous because
of the temperature and entropy discontinuity.

How fast a star is rotating is characterized by different parameters which are appropriate
in different contexts. One of the most frequently used parameter is so-called 'T/W' value
\citep{Tassoul1978book}
which is defined by the ratio of rotational kinetic energy to gravitational energy as
\begin{equation}
	T/W \equiv \frac{\frac{1}{2}\int \rho R^2\Omega^2 dV}{\frac{1}{2}\left|\int \rho\Phi dV\right|}.
\end{equation}
The global parameter is conventionally used to characterize how fast the star is rotating
even when the star is highly differentially rotating. It should be noted that the large degree
of differential rotation does not necessarily means large value of $T/W$. 

We have several parameters to be fixed to compute a model star, which are listed in Table.\ref{table: parameters}
Apart from these we also need the parameters to specify the law and degree of differential rotations
(sec.\ref{sec: rotation laws}) when computing a differentially rotating star.
Physical significance of some of these parameters is worth mentioning. In our models
the core-envelope boundary is characterized by the relative value of pressure to the
central value, $\FP$. It determines the position of the cut-off of the core beyond which
the EOS is switched to that of the finite temperature. Therefore, together with $\rho_{\rm c}$
it determines the core mass and radius. Temperature at the core-envelope boundary,
$\TB$, together with $\rho_{\rm c}$ and $\FP$ fixes the entropy of the envelope,
which determines the mass and the radius of the envelope. $\TB$ does not affect
the core characteristics.
A large value of $\FP$ may result in
a smaller mass of the core than that of the envelope, which
is not natural for a merger remnant model. On the other hand
very small $\FP$ results in a model which is similar
to the completely degenerate model.
In this paper, we adopt the range of $\FP$ with ${\cal O}[\FP]\sim 0.01-0.1$ so that
the models have a sufficiently large core mass but the envelope is also important.

\begin{table}
	\begin{tabular}{cc}\hline
		$\rho_{\rm c}$ & central mass density\\
		$p_{\rm b}/p_{\rm c}\equiv \FP$ & dimensionless pressure at the core-envelope boundary\\
		$\TB$ & temperature at the core-envelope bounday\\
		$(Z/A)_{\rm c}$ & ratio of atomic to baryon number in the core\\
		$X_{\rm H}$ & hydrogen fraction in the envelope\\
		$X_{\rm He}$ & helium fraction in the envelope\\
		$X_{\rm C}$& carbon fraction in the envelope\\ 
		$X_{\rm O}$& oxygen fraction in the envelope\\
		$r_{\rm p}/r_{\rm e}$ & axis ratio\\ \hline
		\end{tabular}
		\caption{Parameters to be fixed to compute our models Apart from these parameters
		we also need to fix the parameters to specify the rotation law in Sec.\ref{sec: rotation laws}
		when dealing with a differentially rotating star.}
		\label{table: parameters}
\end{table}

It should be remarked that \cite{Kadam2016} extended the HSCF scheme to handle two layers
of polytropic gas. Our formulation is different from theirs in the following points. Firstly we 
have the EOS of degenerate electrons in the core and the tabulated finite-temperature EOS
in the envelope, while \cite{Kadam2016} assumes polytropic EOS.
Secondly we allow different temperature at the core-envelope boundary. Thirdly we allow
the stars to have differential rotation, rather than assuming uniform rotation.
All these features in our formulation are motivated by the merger remnant of white dwarf binaries.

\section{Results}
\label{sec: results}

In this paper we assume for simplicity that the core is made of elements whose ratio of mass to atomic
number is $2$. This roughly applies to helium, carbon, oxygen, and neon, For magnesium the ratio is $2.03$,
which is still well approximate by $2$. The simple EOS of completely degenerate electrons is characterized by
the mean molecular weight per electron $\mu_e = Y_e^{-1}$, where $Y_e$ is the electron
fraction. Our core model does not distinguish the difference among heavy elements.

Following \cite{BoshkayevQuevedo2018} we take the central density of the core to be $\rho_{\rm c}<1.37\times 10^{11}{\rm g}{\rm cm}^{-3}$,
beyond which the inverse $\beta$ reaction destabilizes the core.
\footnote{It should be noted that this choice may overestimate the maximum density possible.
Maximum central density of a white dwarf with a typical composition may be less than 
$10^{10}{\rm g}{\rm cm}^{-3}$, beyond which pycnonuclear reaction of carbon occurs in
an C-O white dwarf, while pycnonuclear \citep{Gasques2005}
and electron captures on ${}^{20}{\rm Ne}$ occurs in an O-Ne star \citep{Miyaji1980}.
I thank the anonymous referee for pointing these out and suggesting the appropriate
references.
}

As for the numerical mesh, we use uniformly-separated grid points in $r, \theta$ direction of
the spherical polar coordinate. Since our models are supposed to have the equatorial symmetry,
we restrict $\theta$ as $0\le\theta\le\pi/2$.
All of the models in the paper are computed with the radial grid number. $N_r=200$ and the angular
grid number $N_\theta=100$. Doubling each of the grid numbers results in the residual of order 
$10^{-2}\%$ in the total mass for rapidly rotating (axis ratio = 0.5) models with differential rotation. 
The residuals in gravitational, kinetic and thermal
energy are of the same order. The residual of core mass is larger but at most $1\%$.

%

\subsection{Non-rotating stars}
\begin{figure}
\includegraphics[width=9cm]{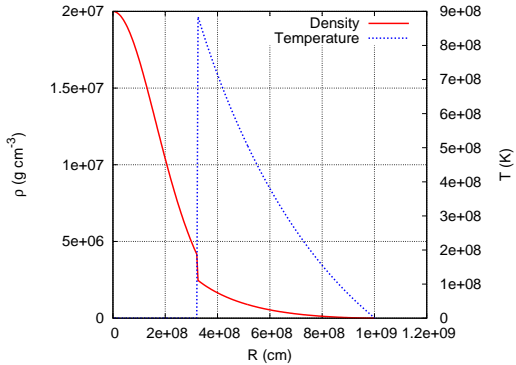}
\caption{Density and temperature distribution of a typical equilibrium state. 
The central density is $\rho_{\rm c} = 2\times 10^7\gcm$ and the temperature at the core-envelope
boundary is $\TB=9\times 10^8$K. The relative pressure at the core-envelope boundary 
and at the center is $\FP=10^{-1}$. It should be noted that due to the finite number of grid
points, the largest temperature shown in the plot seems less than $\TB$, which is an artifact
of the plot.}
\label{fig: rhoTprofile}
\end{figure}

First we see the effect of finite temperature envelope on the globals structure of a star in the non-rotating
limit. In Fig.\ref{fig: rhoTprofile} a typical density and temperature profile is shown as a function of radial coordinate.
The density and the temperature is discontinuous at the core radius $R_c=3.2\times 10^8$cm, though the pressure
is made continuous there (sec.\ref{sec: HSCF}).

\begin{figure}
\includegraphics[width=9cm]{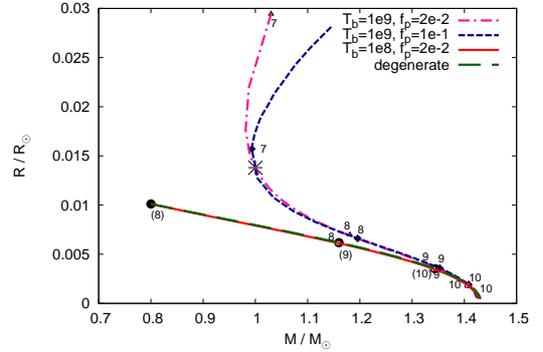}
\caption{Mass-radius relations of non-rotating stars. Mass fraction of elements are
$\XC=0.2$ and $\XO=0.8$. On each curve, the pressure ratio $\FP$ and the
temperature at the core-envelope boundary $\TB$ are fixed.
The numbers attached to the curves are $\log_{10}[\rho_{\rm c}/1~\gcm]$.
The curve labeled as 'degenerate' is the mass-radius relation of the star with 
the same composition. The asterisk on the sequence with $\TB=10^{9}$K and
$\FP=10^{-1}$ marks the point below whose central density the star has
a larger mass in the envelope than in the core. Notice that the sequence
with $\TB=10^8$K and $\FP=2\times 10^{-2}$ is almost on top of the 
completely degenerate one.}
\label{fig: nonrot MR}
\end{figure}

In Fig.\ref{fig: nonrot MR} mass-radius relation of non-rotating stars are plotted. 
Chemical composition of the envelope are $\XC=0.2$ and $\XO=0.8$ in the mass fraction. 
On each curve we change the central
density of the core while fixing the pressure ratio $\FP$ which specifies the boundary between the core and the envelope,
and the temperature $\TB$ there.
For comparison we plot the mass-radius of completely degenerate stars. 
We see the following characteristics of the equilibria. Firstly, as we increase the central
density the core starts to dominate the envelope and the stellar structure tends to resemble
to that of white dwarfs supported by completely degenerate electrons. Deviation
from the degenerate model is seen in the low central density models for sufficiently high
temperature of the envelope. Comparing $\FP=2\times 10^{-2}$ models, we see the lower
temperature ($\TB=10^8$K) sequence is almost on top of the degenerate sequence,
while the higher temperature sequence has the larger radius than the degenerate star
with the same mass. 
As the central density decreases, the core mass decreases while its radius increases. The surface
gravity of the core is reduced and the hot envelope expands rapidly as a function of $\rho_{\rm c}$. 
The mass fraction of the envelope increases and the total mass starts to increase again as the central density decreases.
The increasing envelope mass is supported by the thermal
gas pressure of the hot gas. By increasing the fractional pressure $\FP$ at the core-envelope boundary, we have 
an enhanced effect of hot envelope. 
Since $\FP$ parametrize the cut-off of the core region, 
a smaller value means the core region has a larger fraction in the star. 
As a resutl when $\FP$ decreases, the mass-radius relation of the core asymptotes to that of fully
degenerate stars. The envelope, on the other hand, has a larger entropy and swells up so that
the stellar radius grows, although the mass of the envelope do not contributes to the total mass so much.
When $\FP$ increases, the core mass is reduced and the stellar structure becomes
relatively envelope-dominated. In this case, the envelope can sustain a larger mass
than that of the fully degenerate star.

It should be noted, however, that the models whose envelope mass is
larger than that of the core would not be realized in binary merger events, since the less massive secondary star 
in a binary has a larger radius and is disrupted tidally to accrete onto the more massive primary to form
a hot envelope. On the sequence with $\TB=10^{-9}$K and $\FP=10^{-1}$, the branch above
the 'asterisk' has a larger mass in the envelope than in the core. 

\begin{figure}
\includegraphics[width=9cm]{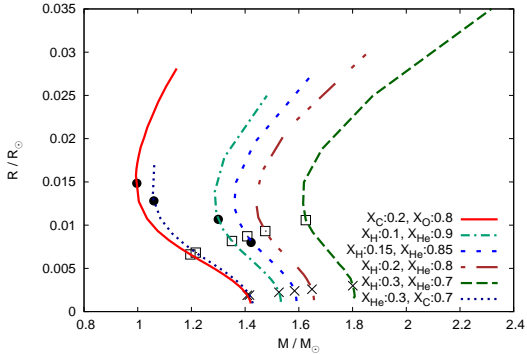}
\caption{Mass radius relations for models with different 
compositions in the envelope. We fix $\TB=10^9$K and $\FP=10^{-1}$.
The range of the central density on these sequence is 
$10^7\le \rho_{\rm c}/1\gcm \le 10^{11}$.
A filled circle marks the point above which the envelope mass exceeds that of the 
core (envelope dominant). The open squares and crosses 
mark the models whose central density are $10^8\gcm$ and $10^{10}\gcm$
respectively.
Dashed curves are always envelope dominant.
}
\label{fig: nonrot composition of envelope}
\end{figure}

We show in Fig.\ref{fig: nonrot composition of envelope} the 
dependence of the mass-radius relation on the
envelope's chemical composition. 
We compare the cases that have
different fraction of $(X_{\rm H}, X_{\rm He}, X_{\rm C}, X_{\rm O})$ in the envelope.
Filled circles mark the points above which the envelope mass exceeds
that of the core. On dashed curves the envelope mass always
exceeds that of the core.
We notice that the difference in the fraction of the hydrogen 
in the envelope affects the relation more than the difference
in the other elements. As we increase the hydrogen fraction
more massive models are allowed. On the other hand the models
with no hydrogen in the envelope differ slightly each other.
This feature is understood as originating from the difference in the mean 
molecular weight. For gas with fixed density and temperature, 
as the mean molecular weight decreases
the pressure increases. The mean molecular weight decreases
as we increase the hydrogen fraction, while it does not change
as for the other elements.
Thus the part of the thermal pressure supporting the envelope mass and radius 
rises as the hydrogen fraction increases. Consequently higher
mass model is realized for the same central density and the 
temperature of the core-envelope boundary when we increase the
hydrogen fraction.

\subsection{Uniformly rotating stars}
Next we summarize the characteristics of uniformly rotating stars.

Our modified HSCF code can compute a sequence of rotating stars up to
the mass-shedding limit at which the matter at the equatorial surface is
in the state of balance between centrifugal and gravitational force. The enthalpy 
gradient does not contributes to the mechanical balance at the limit.
 With a faster rotation the matter would be shed from the
surface of the star, thus the equilibrium sequence terminates at the limit.

\begin{figure}
\begin{center}
\includegraphics[width=9cm]{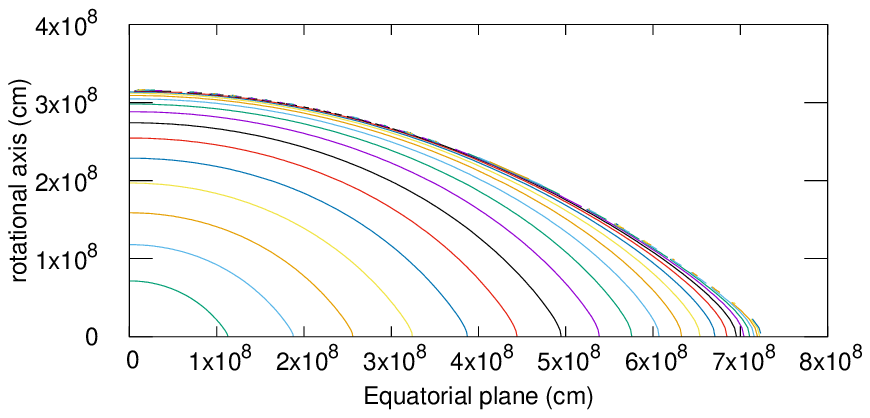}
\caption{Pressure contour of a uniformly rotating star close to its mass-shedding limit. Thirty
levels surfaces are equally spaced in logarithm of pressure. 
Assuming the axial and equatorial symmetry, we only plot a quadrant of the
meridional cross section. 
The central density is $10^8~\gcm$, pressure at the core-envelope
boundary is $\FP=10^{-1}$ and temperature at the boundary is $\TB=5\times 10^8$K.
The total mass of the star is 1.37$M_\odot$, while the core mass is $0.719M_\odot$. 
The equatorial core radius is $2.09\times 10^8{\rm cm}$.
Dimensionless parameter $T/W=2.61\times 10^{-2}$, which corresponds to rotational angular frequency $\Omega=0.687$Hz.
}
\label{fig: unirot logP contour}
\end{center}
\end{figure}

In Fig.\ref{fig: unirot logP contour} logarithm of the pressure is shown
as a contour plot. 
The model is slightly below the mass shedding limit with $T/W=1.78\times 10^{-2}$.
The highly flattened stellar configuration due to the centrifugal force is apparent.

\begin{figure}
\begin{center}
\includegraphics[width=9cm]{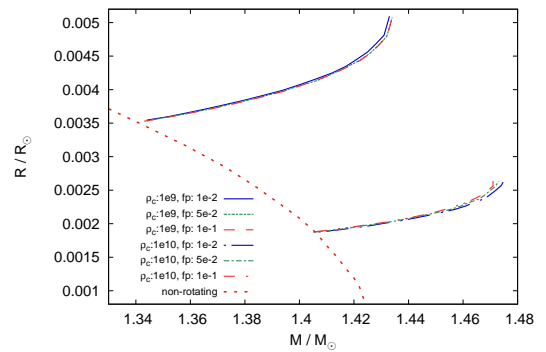}
\caption{Mass-radius relation of uniformly rotating model seque in the envelope.
The rightmost point of each curve corresponds to the mass-shedding limit.
The curve labeled as "non-rotating" is the mass-radius curve for the non-rotating model.
}
\label{fig: MR unirot}
\end{center}
\end{figure}

In Fig.\ref{fig: MR unirot} we plot the mass-radius relation of a uniformly rotating
star's equilibrium sequence up to its mass-shedding limit. All the models have the 
fixed temperature at the core-envelope boundary  $\TB=5\times 10^8$K 
and the composition both in the core and the envelope $(\XC, \XO)=(0.2, 0.8)$.
To construct each sequence we additionally fix $\rho_{\rm c}$ and $\FP$. 
For uniformly rotating stars, both the total and the core mass increase at
most by a few percent even at the mass-shedding limit. The core radius increases
by a few percent while the radius of the envelope increases by tens of percent.
For a uniformly rotating star to have a large mass, it needs to have large $T/W$
so that the rotational contribution to the force balance of the stellar fluid becomes 
significant. When the stellar EOS is soft (i.e. the adiabatic index is small), the
rotational sequence of equilibrium reaches the mass-shedding limit before
$T/W$ reaches such a large value that the mass increases significantly.
The volume-averaged adiabatic index of the envelope is about $1.6$, which is
too soft to support a large increase in mass compared to the non-rotating model.
As is seen in Fig.\ref{fig: MR unirot} the mass-radius relation is scarcely changed 
by changing the pressure fraction $\FP$.

\subsection{Differentially rotating stars: Yoon07 rotation law}
As the first examples of differentially rotating models, we consider stars whose rotational
profile is that of Sec.\ref{sec: rotation law Yoon}. We hereafter call it Yoon07 rotation law.

\begin{figure}
\begin{center}
\includegraphics[width=9cm]{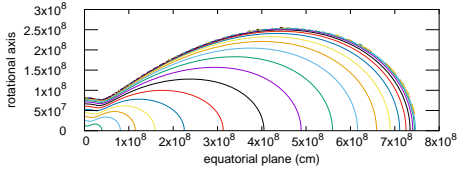}
\caption{Pressure contour of a differentially rotating star with Yoon07 profile close to its mass-shedding limit. Thrity
levels surfaces are equally spaced in logarithm of pressure. 
The central density is $10^8~\gcm$, pressure at the core-envelope
boundary is $\FP=10^{-1}$ and temperature at the boundary is $\TB=5\times 10^8$K.
The total mass of the star is 2.31$M_\odot$, while the core mass is $0.785M_\odot$. 
The equatorial core radius is $1.01\times 10^8{\rm cm}$.
Dimensionless parameter $T/W$ amounts to $6.20\times 10^{-2}$ and the rotational angular frequency $\Omega=0.852$Hz
at its equatorial surface.
Pressure maxima is at the coordinate origin.
}
\label{fig: irot4 logP contour}
\end{center}
\end{figure}

The rotation law allows much larger value of angular momentum and $T/W$ for given
$\rho_{\rm c}, \FP, \TB$ than the corresponding models with uniform rotation. 
In Fig.\ref{fig: irot4 logP contour} we show an  isocontour plot of pressure 
for a typical model of nearly mass-shedding limit that is to be compared with Fig.\ref{fig: unirot logP contour}.
The star has a small core whose equatorial radius is roughly $1/7$ of the envelope radius.
The configuration is highly flattened due to the centrifugal force, with a small flattened core
surrounded by a massive disk-like envelope.

\begin{figure}
\begin{center}
\includegraphics[width=9cm]{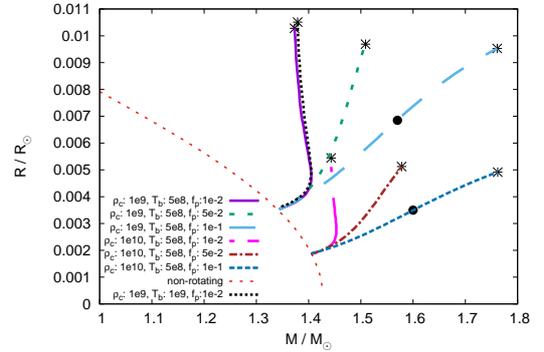}
\caption{Mass and radius of the Yoon07 rotation law
for different values of $\FP$. 
We fix the chemical composition of both in the core and
in the envelope as $(\XC, \XO)=(0.2, 0.8)$. Filled circles mark
the points above which the envelope mass exceeds that
the core. The asterisks show the mass-shedding limit of
the sequences.
}
\label{fig: MR yoon}
\end{center}
\end{figure}
Mass-Radius relations (Fig.\ref{fig: MR yoon}) show that a 
differentially rotating sequence may reach larger mass and radius
models than the uniformly rotating counterpart.
%
%
For given $\rho_{\rm c}$ and $\TB$, larger $\FP$ leads to larger increments
in mass and radius compared with the uniformly rotating model seen in Fig.\ref{fig: MR unirot}. 
This is because a larger $\FP$ means the cold degenerate core is truncated at a smaller
mass and the hot envelope is attached to it with the larger pressure.
It corresponds to a smaller contribution to the
entire star from the core which rotates slowly. The hot envelope
may be rapidly rotating in this rotation law to sustain a large mass and to be extended
to a large radius.

\begin{figure}
   \includegraphics[width=9cm]{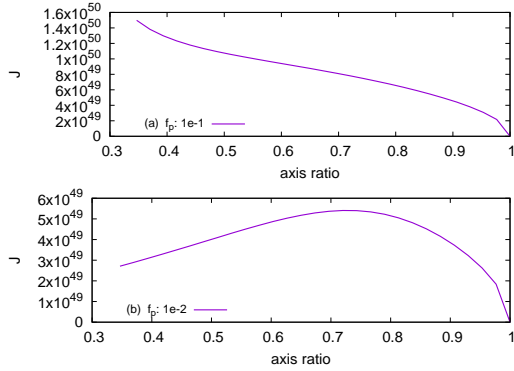}
   \caption{Angular momentum as a function of axis ratio. We fix $\TB=5\times 10^8$K
   , $\rho_c=10^9\gcm$, and $(\XC, \XO)=(0.2, 0.8)$. 
   The angular momentum is unit of g${\rm cm}^2 {\rm s}^{-1}$.
   The top panel (a) is  for $\FP=10^{-1}$, 
   while the bottom one (b) is for $\FP=10^{-2}$.
   For non-rotating stars the axis ratio is unity.}
	\label{fig: J yoon}
\end{figure}

A qualitatively different behaviour of sequences for 
different $\FP$ is noticed in Fig.\ref{fig: MR yoon}. 
For instance, the sequences with $\FP=10^{-1}$ and $5\times 10^{-2}$ show 
monotonic increase in radius and mass as the star spins up, while the sequence
with $\FP=10^{-2}$ shows non-monotonic change in mass.
The difference originates from our parametrization to construct the sequences.
As for the parameter to quantify the rotational deformation of a star 
we use the axis ratio which is the ratio between the polar and the equatorial radii
of the star. For a star with sufficient mass contribution from its envelope
whose $\FP$ value is large, the deformation quantified by the axis ratio
monotonically corresponds to other characteristics of measuring stellar rotation
rate such as angular momentum or $T/W$ parameter. In Fig.\ref{fig: J yoon} (a) 
the angular momentum for $\FP=10^{-1}$ case is plotted as a function of
the axis ratio. It should be noted for non-rotating stars the axis ratio is unity. 
The angular momentum increases as the stellar deformation by
rotation is enhanced. Therefore the parametrization by the axis ratio
is faithful to the rapidness of rotation measured by the angular momentum.
 For $\FP=10^{-2}$ case (panel (b)),
however, the angular momentum is not a monotonic function of the axis ratio. Thus
the sequence is not regarded as that of increasing degree of rotation.

\subsection{Differentially rotating stars: Kepler rotation law}
\begin{figure}
\begin{center}
\includegraphics[width=9cm]{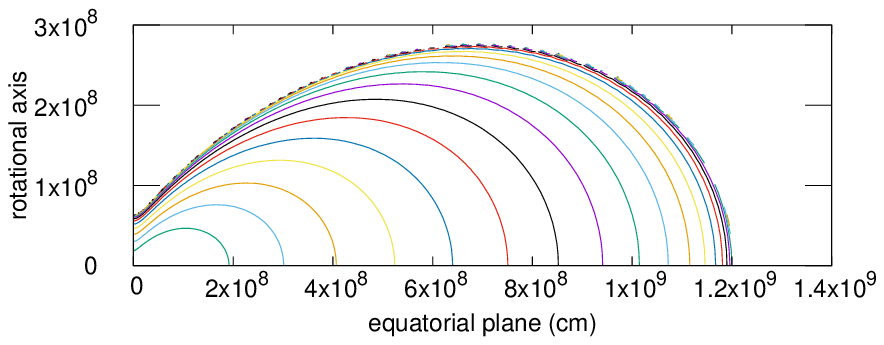}
\caption{Pressure contour of a star close to its mass-shedding limit for Kepler rotation law. Thirty
levels surfaces are equally spaced in logarithm of pressure. 
Assuming the axial and equatorial symmetry, we only plot a quadrant of the
meridional cross section. 
The central density is $10^8~\gcm$, pressure at the core-envelope
boundary is $\FP=10^{-1}$ and temperature at the boundary is $\TB=5\times 10^8$K.
The total mass of the star is 3.15$M_\odot$, while the core mass is $1.56M_\odot$. 
The equatorial core radius is $3.33\times 10^8{\rm cm}$.
Dimensionless parameter $T/W$ is $2.11\times 10^{-1}$ and the rotational angular frequency $\Omega=0.488$Hz
at its equatorial surface.
}
\label{fig: irot3 logP contour}
\end{center}
\end{figure}

Next we see the results for the rotation law in Sec.\ref{sec: rotation law Kepler}.
We call the law as Kepler rotation.
A contour plot of a nearly mass-shedding stellar model with Kepler rotation law
is given in Fig.\ref{fig: irot3 logP contour}. The core is rotationally deformed in
a hamburger shape, on top of which a torus-like envelope resides.
The pressure contour is highly elongated in the equatorial plane due to strong centrifugal
force from the differential rotation. Compared with a nearly mass-shedding case of 
Yoon07 (Fig.\ref{fig: irot4 logP contour}) the envelope is 
expanded more in the Kepler rotation law model. 

\begin{figure}
	\includegraphics[width=9cm]{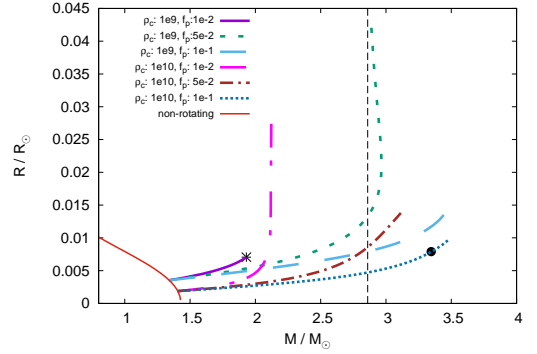}
	\caption{Mass-radius relation for the "Kepler." rotation law.
	All the models have $\TB=5\times 10^8$K and $(\XC, \XO)=(0.2, 0.8)$.  
	The filled circle marks the point above which the envelope mass exceeds that of the core,
	while the asterisk marks the mass-shedding limit. The dashed line corresponds
	to the mass twice as large as the Chandrasekhar mass of the non-rotating model
	for this composition ($M=1.43M_\odot$).}
	\label{fig: MR kepler}
	\end{figure}

We plot the mass-radius relation for
$\TB=5\times 10^8$K and $(\XC, \XO)=(0.2, 0.8)$ in Fig.\ref{fig: MR kepler}.
The vertical dashed line corresponds to twice the Chandrasekhar mass of a cold non-rotating
white dwarf. Beyond the line a sequence may lose its meaning for a model of a merger
remnant. Numerical computations of sequences without asterisk are stopped
	at axis ratio equals 0.2. The sequence may be extended beyond
	this point and the stellar structure may take
	a "doughnuts" shape with off-center density maximum. Since our 
	current code does not converge for very small axis ratio case,
	we stop the computation there.
\begin{figure}
	\includegraphics[width=9cm]{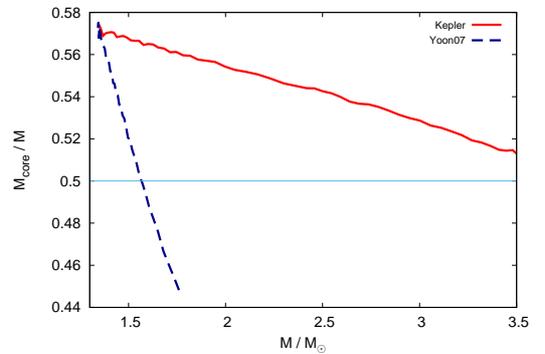}
	\caption{Ratio of the core mass to the total mass for Yoon07 and
	Kepler rotation laws. At the top-left point the model is non-rotating
	and the axis ratio is decreased to have models with larger degree of rotation
	along the sequence.
	Other parameters are fixed as $\rho_c = 10^9\gcm$
	, $\TB=5\times 10^8$K, $\FP=10^{-1}$ 
	and $(\XC, \XO)=(0.2, 0.8$).}
	\label{fig: Mcore ratio}
	\end{figure}
The mass and the $T/W$ parameter 
may become also larger for the Kepler law. It should be noted that the core mass
is also higher for the Kepler case as is seen in Fig.\ref{fig: Mcore ratio}. For given
composition, $\TB$ and $\rho_c$, the model with Kepler rotation has much
higher mass for a given core mass ratio, which means both core and total mass
is much higher for Kepler rotation models.
This is because the Yoon07
profile has a slowly rotating core while the Kepler has a rapidly rotating one.
The centrifugal support against the self-gravity is larger for the latter case,
which enables the larger core mass.

\section{Some equilibrium sequences of simplified evolution}
As for evolutions and final fates of merger remnants of white dwarf binaries, there
have been theoretical studies focusing on different phases of the remnant
evolutions. Three-dimensional numerical hydrodynamic codes developed by different research
groups have been used to follow successive phases of binary-inspirals, mergers and formation of remants
\citep{Guillochon2010,Dan2011,Dan2012,Dan2014,Tanikawa2015,Sato2015,Sato2016}.
These studies mainly investigate the possibility of explosive carbon burning that
leads to prompt explosion scenario of type Ia supernovae. \cite{Dan2014} and \cite{Sato2015}
systematically surveyed the parameter space of possible progenitors of the supernovae.
The other studies \citep{Yoon2007,Schwab2012,Shen2012} have focused on evolutions
of remnants in much longer time scale. In these studies outcomes of three-dimensional 
hydrodynamical simulations are projected onto one-dimensional codes that solve
thermal, chemical and rotational evolution of a remnant
in quasi-hydrostatic approximation. The secular evolution of remnants
toward the so-called 'delayed explosion' is the main issue to be investigated.
On the other hand \cite{Zhang2014} studied evolution of merger remnants
as progenitors of R Coronae Borealis and extreme helium stars.

As an application of our new code, we construct equilibrium sequences
that may mimic early evolutionary paths of merger remnants.
Our motivation here is not to construct elaborated models of progenitors
of supernovae or carbon rich subdwarfs. We study some
characteristics of sequences and check viability of the rotating equilibria
by a simple argument.
Thus we do not go into a detailed study including thermal relaxation, 
nuclear reactions and radiative cooling, but study constrained sequences
of equilibria with some global parameters being fixed. 
Particularly it should be noted that the temperature at the
bottom of the envelope $\TB$ may exceed $10^8$K for
some models below. In those cases helium burning may contribute
to the heating of the envelope and the simple approximation
here may not be adequate. The effect of the heating is beyond
our scope here.

%
\subsection{Sequences with fixed entropy in the envelope}
First we fix specific entropy $S_{\rm b}$ of the hot envelope as well as the masses
of the core and the entire star. The sequences are intended to mimic evolutions through
rapid angular momentum loss compared with the thermal relaxation and dissipation.
We do not specify the mechanism of angular momentum loss
and assume the rotation law to be unchanged during an evolution.
The sequences here are computed by iteratively solving for $\rho_{\rm c}$,
$f_{\rm p}$, and the axis ratio with the total mass $M$, the core mass $M_{\rm core}$
, and the specific entropy of the envelope $S_{\rm b}$ being fixed.

\begin{figure}
	\centering
	\includegraphics[width=9cm]{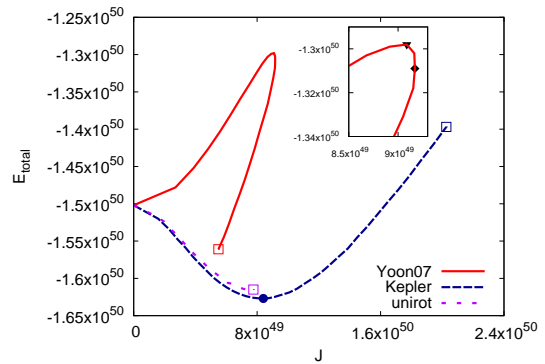}
	\caption{Sequences of equilibrium models on which the total mass 
	$M=1.16M_\odot$,
	the core mass $M_{\rm core}=0.84M_\odot$ 
	and the specific entropy $S_{\rm b}=1.93\times 10^8~({\rm erg}~{\rm g}^{-1}{\rm K}^{-1})$ are fixed. 
	The entropy value corresponds to $\TB=1\times 10^8$(K) for the non-rotating model.
	Total energy of a star $E_{\rm total}$
	(in ergs) is plotted against the total angular 
	momentum $J$ (in ${\rm g}{\rm cm}^2{\rm s}^{-1}$).
	The open squares mark the mass-shedding limits of the sequence.
	The local minima of $E_{\rm total}$ are marked by the filled circles.
	The key "unirot" means the sequence with the uniform rotation.
	An enlargement of the part of Yoon07 sequence is shown in the inset.
	The filled diamond corresponds to the maximum of $J$, while the 
	filled triangle corresponds to the maximum of $E_{\rm total}$.
	The composition of the envelope is $X_{\rm He}=1$.
	}
		\label{fig: M-Sb-const}
	\end{figure}
In Fig.\ref{fig: M-Sb-const} total energy $E_{\rm total}$ of stars is plotted as a function of
angular momentum $J$. The total energy is defined as
\begin{equation}
	E_{\rm total} = \frac{1}{2}\int\rho\Phi dV + \frac{1}{2}\int\rho(\Omega r\sin\theta)^2 dV
	+ \int\rho u dV,
\end{equation}
 	where $u$ is specific internal energy. Each term corresponds to gravitational, kinetic and
	thermal energy, respectively.
The sequences have total mass $M=1.16M_\odot$ and the core mass $M_{\rm core}=0.84M_\odot$.
As for the composition of the envelope, we follow the scheme in \citet{Marsh2011}, whose Fig.1
shows the possible combinations of white dwarf's composition in the parameter space
of the binary component's masses. For Fig.\label{fig:  M-Sb-const}, the lighter
component of the progenitor ($M=0.32M\odot$)  consists of pure
helium ($X_{\rm He}=1$) for simplicity. We do not consider nuclear reactions to change
the composition further. 

\begin{figure}
	\centering
	\includegraphics[width=9cm]{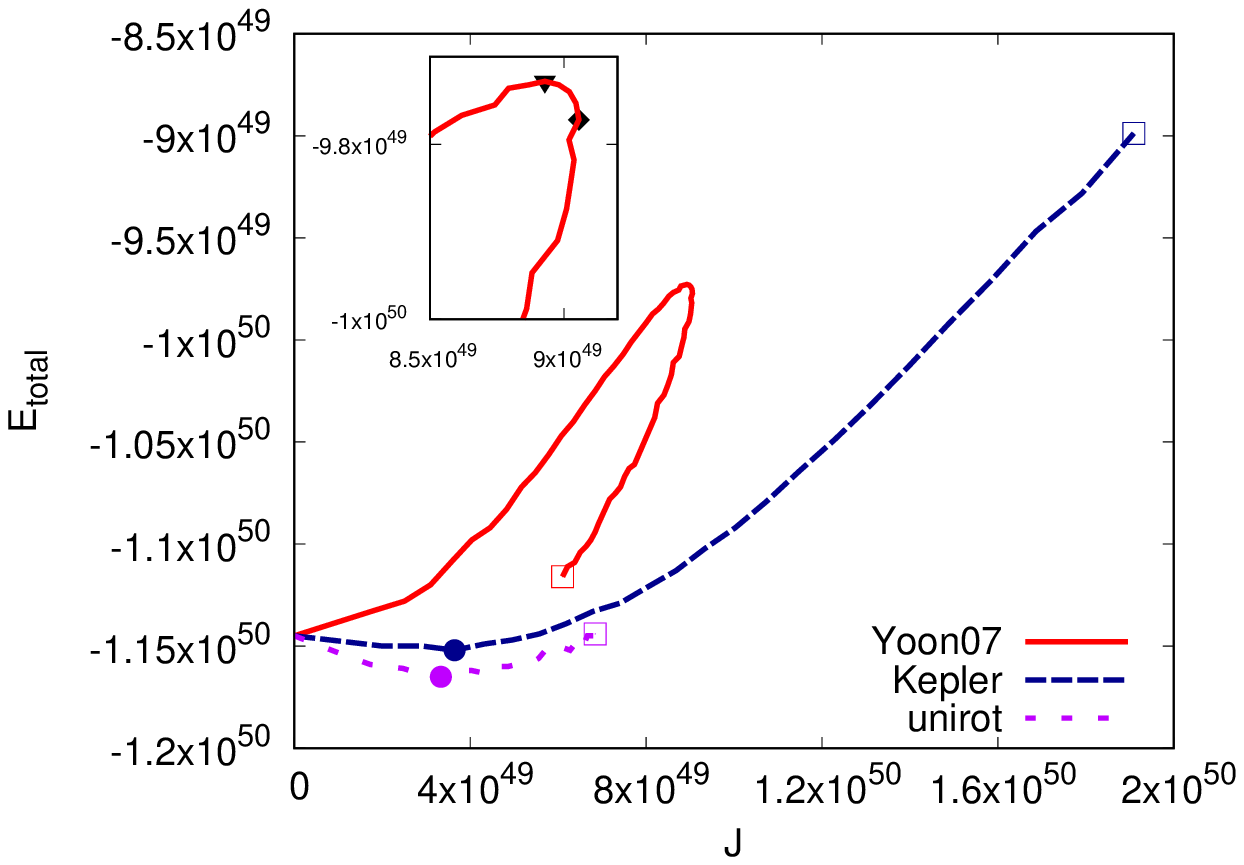}
	\caption{Same as Fig.\ref{fig: M-Sb-const} except the total mass 
	$M=1.0M_\odot$,
	the core mass $M_{\rm core}=0.7M_\odot$ 
	and the specific entropy $S_{\rm b}=3.11\times 10^8~({\rm erg}~{\rm g}^{-1}{\rm K}^{-1})$ are fixed. 
	The entropy value corresponds to $\TB=5\times 10^8$(K) for the non-rotating model.
	Total energy of a star $E_{\rm total}$
	(in ergs) is plotted against the total angular 
	momentum $J$ (in ${\rm g}{\rm cm}^2{\rm s}^{-1}$).
	The open squares mark the mass-shedding limits of the sequence.
	The local minima of $E_{\rm total}$ are marked by the filled circles.
	An enlargement of the part of Yoon07 sequence is shown in the inset.
	The filled diamond corresponds to the maximum of $J$, while the 
	filled triangle corresponds to the maximum of $E_{\rm total}$.
	The composition of the envelope is $X_{\rm He}=1$.
	}
		\label{fig: M and Sb const Tb5e8}
	\end{figure}

\begin{figure}
	\centering
	\includegraphics[width=9cm]{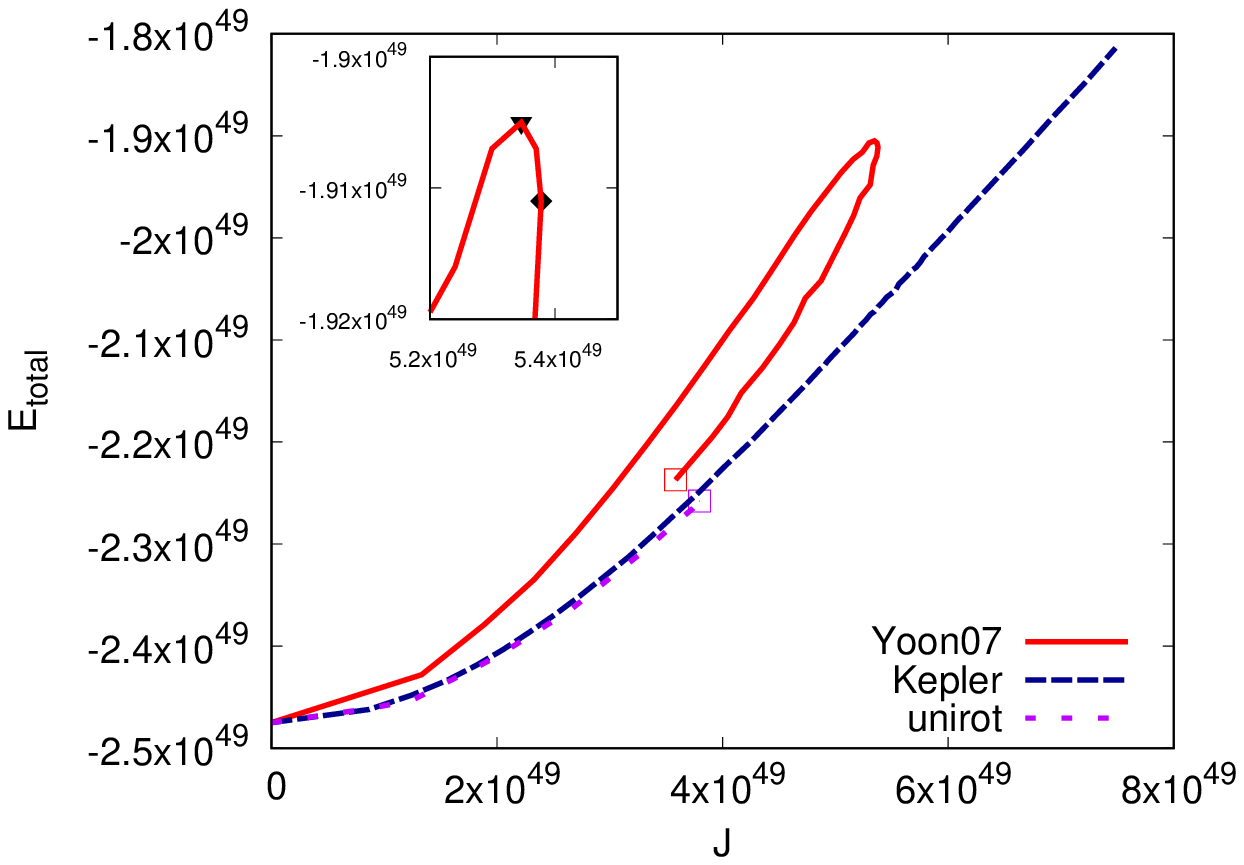}
	\caption{Same as Fig.\ref{fig: M-Sb-const} except that the total mass 
	$M=0.5M_\odot$,
	the core mass $M_{\rm core}=0.3M_\odot$ 
	and the specific entropy $S_{\rm b}=2.82\times 10^8~({\rm erg}~{\rm g}^{-1}{\rm K}^{-1})$ are fixed. 
	The entropy value corresponds to $\TB=1\times 10^8$(K) for the non-rotating model.
	Total energy of a star $E_{\rm total}$
	(in ergs) is plotted against the total angular 
	momentum $J$ (in ${\rm g}{\rm cm}^2{\rm s}^{-1}$).
	The open squares mark the mass-shedding limits of the sequence.
	These models are regarded as remnants of He+He white dwarf binaries.	
	The composition of the envelope is $X_{\rm He}=1$.}
		\label{fig: M0.5 and Sb const}
	\end{figure}

\begin{figure}
	\centering
	\includegraphics[width=9cm]{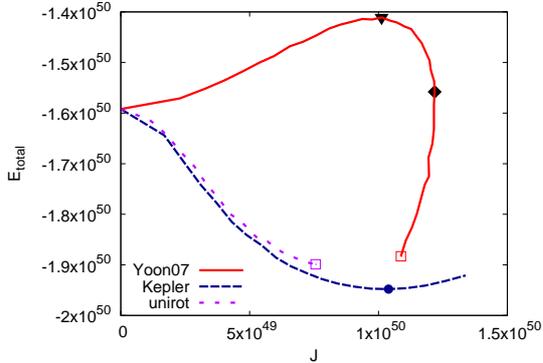}
	\caption{Same as Fig.\ref{fig: M-Sb-const} except that the total mass 
	$M=1.26M_\odot$,
	the core mass $M_{\rm core}=0.72M_\odot$ with the envelope composition
	of $\XC=0.6$, $\XO=0.4$.
	The specific entropy $S_{\rm b}=6.16\times 10^7~({\rm erg}~{\rm g}^{-1}{\rm K}^{-1})$ are fixed. 
	The entropy value corresponds to $\TB=1\times 10^8$(K) for the non-rotating model.
	Total energy of a star $E_{\rm total}$
	(in ergs) is plotted against the total angular 
	momentum $J$ (in ${\rm g}{\rm cm}^2{\rm s}^{-1}$).
	The open squares mark the mass-shedding limits of the sequence.
	The filled diamond corresponds to the maximum of $J$, while the 
	filled triangle corresponds to the maximum of $E_{\rm total}$.
	These models are regarded as remnants of CO+CO white dwarf binaries.	
	}
		\label{fig: M and Sb const XC0.6 XO0.4}
	\end{figure}

On these sequences physically allowed segments are the ones on which 
total energy decreases as the total angular momentum decreases.
Thus the natural branch that models an evolutionary path of a star should satisfy,
\begin{equation}
	\left.\frac{dE_{\rm total}}{dJ}\right|_{S_{\rm envelope}}>0,
	\label{eq: condition of M-Tb sequences}
\end{equation}
where $S_{\rm envelope}$ is the total entropy of the envelope. For the current isentropic
envelope $S_{\rm envelope}=M_{\rm envelope} S_{\rm b}$.
For the Kepler sequence, the total energy has
a minimum marked by the filled circle. The segment on the right of the minima
are the allowed ones if a star is not spun up by some external torque (e.g. due
to accretion of fall back debris). Notice that the open squares mark the mass-shedding
limit. A rapidly rotating star may evolve along
the segment by losing their angular momentum and energy until they hit the minimum
of $E_{\rm total}$. After that the star may not evolve along the adiabatic sequence.
On the other hand the uniform rotation sequence (whose key is "unirot") has no extremum, 
but terminates at the mass-shedding.
Thus the uniformly rotating star with the parameter set may not be a model of the
adiabatic spin-down evolution.
Another curious feature is that the uniformly rotating sequence and the Kepler
sequence are on top of each other. Similar behaviour is seen in the figures
that follow (Fig.\ref{fig: M and Sb const Tb5e8} - Fig.\ref{fig: M and J const COCO}).
A Kepler model has a uniformly rotating core which has a dominant
contribution to angular momentum when the total angular momentum is as small
as that of mass-shedding model of uniformly rotating star. On the other hand
the total energy is
dominated by the gravitational binding energy which is mainly controlled by
the massive core. As a result the uniformly rotating sequence and the 
Kepler sequence are close to each other. Beyond the angular momentum of 
the mass-shedding limit of the uniformly rotating sequence, 
the contribution from the differentially rotating envelope 
is no longer neglected on the Kepler sequences.

For Yoon07 rotation law, the situation is completely different. The sequence 
has two allowed segments. As is seen in the inset, the sequence has a maxima
of total energy (marked by a filled triangle) and a maxima of angular momentum
(marked by a filled diamond). The upper of the allowed segments starts from the maxima
of $E_{\rm total}$ and terminates at the non-rotating star. The dimensionless
rotational parameter $T/W$ spans $0\le T/W\le 0.068$.
The lower segment starts from 
the maxima of $J$ and terminates at the mass-shedding limit. 
Along it $T/W$ decreases from $0.068$ to $0.024$ (at the mass-shedding).
The latter segment is a non-trivial one appearing as a manifestation of high degree
of differential rotation. 

For higher entropy (temperature) of the envelope the uniformly rotating sequence may have
a minimum of $E_{\rm total}$. In Fig.\ref{fig: M and Sb const Tb5e8} we plot model sequences
with the higher entropy for the similar mass as in Fig.\ref{fig: M-Sb-const}. The composition
of the envelope is the same. Now the uniformly rotating sequence has a minima of $E_{\rm total}$
and there is a short segment from the mass-shedding limit (open square) to the minimum
which satisfies Eq.(\ref{eq: condition of M-Tb sequences}). 
We see the qualitatively similar behaviour for Yoon07 and Kepler 
sequences as in Fig.\ref{fig: M-Sb-const}.

In Fig.\ref{fig: M0.5 and Sb const} we see sequences of less massive remnants
with $M=0.5M_\odot$ and $M_{\rm core}=0.3M_\odot$. The composition
of the envelope is $X_{\rm He}=1$, thus the sequence may be regarded as remnants
of He+He white dwarf binaries. Qualitative behaviour of the sequence
for Yoon07 is similar to that in Fig.\ref{fig: M-Sb-const}. As for Kepler and
uniform rotation sequences, there are no minima on the sequence thus the model
stars may spin down to non-rotating limit if the angular momentum loss takes
place sufficiently fast. 
Notice that the rightmost
point of the Kepler sequence is not a mass-shedding limit.
It is the point beyond which our code gives no
convergence, thus we terminated to construct a model.

In Fig.\ref{fig: M and Sb const XC0.6 XO0.4} we see sequences of remnants with
more massive envelope, for which $M=1.26M_\odot$ and $M_{\rm core}=0.72M_\odot$. 
As we have envelope of $M_{\rm envelope}=0.54M_\odot$, we choose the composition
of it as carbon and oxygen \citep{Marsh2011} and assume $\XC=0.6$ and $\XO=0.4$.
Thus the sequence is regarded as remnants
of CO+CO white dwarf binaries. 
Qualitative behaviour of the sequence
for Yoon07 is similar to that in Fig.\ref{fig: M-Sb-const}. 
It has two definite segments corresponding to
different rapid spin-down paths.
As for uniform rotation sequence, 
the minimum of $E_{\rm total}$ is at the mass-shedding, thus the sequence is not
regarded as a rapid spin-down evolution as in Fig.\ref{fig: M-Sb-const}. 
The Kepler sequence has a minimum
of $E_{\rm total}$, the right segment of which is regarded
as a rapid spin-down path of evolution. Notice that the rightmost
point of the Kepler sequence is not a mass-shedding limit
but merely the point beyond which our code gives no
convergence.

%
\subsection{Sequences with fixed angular momentum}
We construct sequences by fixing angular momentum and total mass
of stars as well their core mass.
These sequences are intended to mimic the case when
radiative cooling of the envelope is so efficient that stars
do not spin down during the evolution.
The sequences here are computed by iteratively solving for $\rho_{\rm c}$,
$\TB$, and the axis ratio, with the total mass $M$, the core mass $M_{\rm core}$
, and the total angular momentum $J$ being constrained.
On these sequence a natural evolutionary path is their segments
on which the total energy $E_{\rm total}$ decreases as the
entropy of their envelope decreases,
\begin{equation}
	\left.\frac{dE_{\rm total}}{dS_{\rm b}}\right|_{J}>0,
	\label{eq: condition of M-J sequences}
\end{equation}
When the condition is satisfied, a star cools down along 
the sequence by conserving angular
momentum until it hits a minimum of $E_{\rm total}$. 
After that the angular momentum loss needs to be taken into
account to consider the evolutionary path of the remnant.

\begin{figure}
	\centering
	\includegraphics[width=9cm]{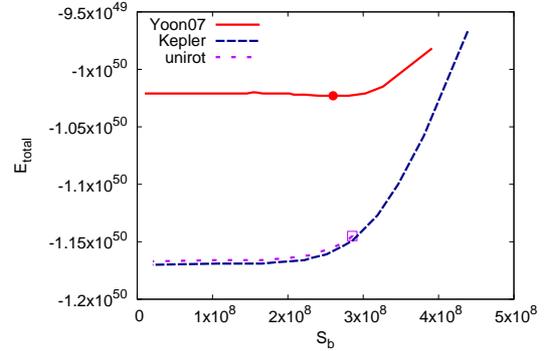}
	\caption{Sequences of equilibrium models on which the total mass
	the total mass $M=1.0M_\odot$,
	the core mass $M_{\rm core}=0.7M_\odot$ 
	and the angular momentum $J=7.1\times 10^{49} {\rm g}{\rm cm}^2{\rm s}^{-1}$ 
	are fixed. The horizontal axis is the specific entropy of the envelope
	$S_{\rm b}$ in ${\rm erg}~{\rm g}^{-1}{\rm K}^{-1}$. The vertical axis
	is the total energy of stars $E_{\rm total}$ in ergs.
	The open square marks the mass-shedding limits of the sequence.
	The local minimum of $E_{\rm total}$ is marked by the filled circle.
	The mass-shedding model for the uniform rotation law has
	$\TB=2.7\times 10^8$K, while the leftmost points of the Yoon07
	and Kepler sequences have $\TB=7\times 10^8$K.
	The mass-shedding model for the sequence with uniform rotation law has
	$\TB=2.7\times 10^8$K while the left most point.
	of it has $\TB=1\times 10^6$K.
	The composition of the envelope is $X_{\rm He}=1$.
		}
		\label{fig: M and J const}
	\end{figure}
In Fig.\ref{fig: M and J const} we plot $E_{\rm total}$ as functions of
$S_{\rm b}$ for different rotation laws. Here the total mass and the core
mass is fixed as $M=1.0M_\odot$ and $M_{\rm core}=0.7$. The envelope
composition is fixed as $\XHe=1$. We fix the angular momentum of all the
sequences to be $J=7.1\times 10^{49}{\rm g}{\rm cm}^2{\rm s}^{-1}$,
which is the angular momentum of the uniformly rotating model at its mass-shedding
limit (open square) for $\TB=2.7\times 10^8$K.
\footnote{Notice that this may underestimate the angular momentum of
remnants which are supposed to be differentially rotating. More realistic
value of angular momentum may be several factors or an order of magnitude larger,
thus it may be of the order of $10^{50}{\rm g}{\rm cm}^2{\rm s}^{-1}$ \citep{Gourgouliatos_Jeffery2006}.
}
The uniformly rotating sequence and the Kepler sequence
with the same angular momentum satisfies Eq.(\ref{eq: condition of M-J sequences})
in the entropy range shown here. Thus a remnant formed on these sequence
may cool all the way down to $\TB\sim 10^6$K. Yoon07 sequence, on the
other hand, have a minimum. Thus the evolutionary sequence of rapid cooling is terminated
there ($\TB\sim 1.5\times 10^8$K).

\begin{figure}
	\centering
	\includegraphics[width=9cm]{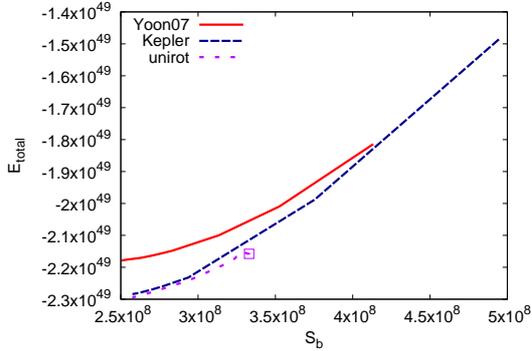}
	\caption{Same as Fig.\ref{fig: M and J const} except that
	$M=0.5M_\odot$,
	the core mass $M_{\rm core}=0.3M_\odot$ 
	and the angular momentum $J=3.5\times 10^{49} {\rm g}{\rm cm}^2{\rm s}^{-1}$ 
	are fixed. The horizontal axis is the specific entropy of the envelope
	$S_{\rm b}$ in ${\rm erg}~{\rm g}^{-1}{\rm K}^{-1}$. The vertical axis
	is the total energy of stars $E_{\rm total}$ in ergs.
	The open square marks the mass-shedding limits of the sequence.
	The local minimum of $E_{\rm total}$ is marked by the filled circle.
	The mass-shedding model for the uniform rotation law has
	$\TB=1.6\times 10^8$K while the leftmost point of it
	has $\TB=5.5\times 10^7$K. For the Yoon07 sequence,  temperature of 
	the envelope at the rightmost and the leftmost
	points are $2.5\times 10^8$K and $5\times 10^7$K respectively.
	For the Kepler sequence, they are $2.5\times 10^8$K and $5.5\times 10^7$K.}
		\label{fig: M0.5 and J const}
	\end{figure}
In Fig.\ref{fig: M0.5 and J const} we plot sequences of fixed angular momentum 
for less massive remnants with $M=0.5M_\odot$ and $N_{\rm core}=0.3M_\odot$.
The fixed angular momentum is that of the mass-shedding limit star for uniform
rotation and $\TB=1.6\times 10^8$K. The sequences here satisfy 
Eq.(\ref{eq: condition of M-J sequences}) and it may be regarded as a rapid cooling
evolutionary paths. Notice that the leftmost and rightmost points of the Yoon07
and the Kepler sequences are points at which our code do not give convergence.

\begin{figure}
	\centering
	\includegraphics[width=9cm]{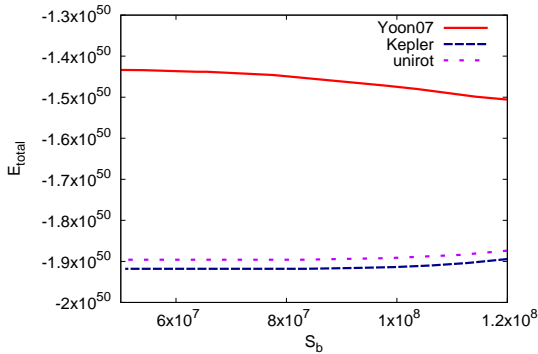}
	\caption{Same as Fig.\ref{fig: M and J const} except that the total mass 
	$M=1.26M_\odot$,
	the core mass $M_{\rm core}=0.72M_\odot$ 
	and the angular momentum $J=7.1\times 10^{49} {\rm g}{\rm cm}^2{\rm s}^{-1}$ 
	are fixed. The composition of the envelope is $(X_{\rm C}, X_{\rm O})=(0.6,0.4)$.
	 For the uniformly rotating sequence, the rightmost and the leftmost
	 points have $\TB=8\times 10^8$K and $7\times 10^6$K.
	 For Yoon07 sequence,  the temperature at the rightmost and the leftmost
	points are $9\times 10^8$K and $4\times 10^7$K respectively.
	For the Kepler sequence, they are $8\times 10^8$K and $3\times 10^7$K.
		}
		\label{fig: M and J const COCO}
	\end{figure}
For more massive remnant of CO+CO white dwarf mergers, the rapid cooling
path may be marginally realized. In Fig.\ref{fig: M and J const COCO}
we plot the models with $M=1.26M_\odot$ and $M_{\rm core}=0.72M_\odot$
whose envelope composition is $\XC=0.6, \XO=0.4$. The Yoon07
sequence does not satisfies Eq.(\ref{eq: condition of M-J sequences}).
The uniform rotation and the Kepler sequences satisfies it
down to $S_{\rm b}\sim 10^8 {\rm erg}~{\rm g}^{-1}{\rm K}^{-1}$, below
which the left hand side of Eq.(\ref{eq: condition of M-J sequences}) is nearly zero. 
The temperature range of the figure is
roughly $\TB=5\times 10^7$K to $3\times 10^8$K, outside of which 
out code do not convergence.

\section{Summary}
We present a new formalism to compute equilibrium of rotating star whose structure
consists of two layers of thermal characteristics. The equation of state of the dense core
is that of completely degenerate electrons, while the equation of state of the hot
envelope is the tabulated 'Helmholtz' equation of state of \cite{Timmes-Swesty2000}.
Our model is meant to mimic remnant objects of binary white dwarf mergers driven
by gravitational radiation. 
In the final phase of in-spiraling the secondary star in a binary fills its Roche lobe and 
dynamically accreted onto the primary. The accreted matter is heated up to form 
a hot envelope, while the primary white dwarf stays in degenerate state as long
as the thermal relaxation is negligible.

Given an angular frequency profile (rotation law), 
our model is parametrized by the composition of its core and envelope, 
the central density $\rho_c$, the temperature at the core-envelope boundary $\TB$, 
and the ratio of pressure at the core-envelope
boundary to the central pressure $\FP$.

When its central density of the core is large enough, a model is close to 
completely degenerate star. The effect of hot envelope becomes significant
when the central density and the core mass decreases. In this case,
thermal pressure of hot envelope supports its mass against the gravity
of the core and itself. The mass-radius relation depends on the chemical
composition of the envelope, especially that of hydrogen. More hydrogen
leads to reduction of the mean molecular weight of the envelope gas,
resulting in the increase of thermal pressure for given temperature.
As a result, the mass and radius are larger compared to the models
lacking hydrogen in the envelope. It should be remarked, however,
a model whose mass in the envelope exceeds that of the core
may not be astrophysically relevant at least as a model of 
a merger remnant of a white dwarf binary. This is because the
hot envelope originates from the secondary star of the binary.
Non-rotating model would not produce super-Chandrasekhar
object even with the presence of hot envelope.

Next we examine the effect of rotation on our models.
Uniform rotation increases the mass and radius of the model 
by at most a few per cent even in the mass-shedding limit.

Introduction of differential rotation changes the picture drastically.
We consider two rotational angular velocity profiles which are motivated by the
preceding studies of merger remnants. One of these profiles
has a slowly rotating core with a rapidly rotating envelope ("Yoon07")
while the other has a rapidly and uniformly rotating core with
a Kepler-like distribution in the envelope ("Kepler").
In both cases the increment of mass and radius are more than
the uniformly rotating counterparts. Especially in the Kepler
case, highly super-Chandrasekhar models are possible whose
mass doubles the Chandrasekhar mass of the non-rotating
star, without encountering the dominance of envelope mass.

Another traits of differentially rotating sequences are the importance
of $\FP$ parameter. $\FP$ is the parameter which determines where the
core is truncated and the envelope is attached. Larger $\FP$ means
the bottom of the envelope have larger pressure, which makes the larger
mass of the envelope possible. As a consequence, a larger $\FP$
model has a larger total mass. This is not apparent in the uniformly
rotating cases, where mass-shedding limit is reached before the effect
of $\FP$ becomes important. For differentially rotating stars, the 
mass-radius relation is strongly affected by $\FP$.

As applications to the early phase of the evolution of the merger remnants,
we consider sequences with total mass and the core mass (therefore the
envelope mass as well) being fixed. 
When we fix the specific entropy
of the envelope $S_{\rm b}$, we may reduce the total angular momentum
of the stars to construct the rapidly spinning-down sequences.
As the angular momentum decreases, the total energy should
not increase if the sequence is regarded as an evolutionary path.
We see the sequences of uniform rotation and of Kepler rotation law
are similar although the latter is extended to much larger value
of angular momentum due to the strong differential rotation.
For a massive remnant with $M\sim 1M_\odot$ and with helium envelope, 
both of the sequences have minima of total energy when the entropy
is large enough (Fig.\ref{fig: M and Sb const Tb5e8}). Thus a rapid spin-down evolution
of these models are possible down to a finite value of angular momentum.
Beyond that the thermal relaxation should not be neglected. For a smaller
entropy envelope, however, the uniformly rotating star do not have an energy
minimum (see Fig.\ref{fig: M-Sb-const}) and the sequence may not
be regarded as that of the rapid spin-down. For
a less massive sequence with $M=0.5M_\odot$, the minima of the energy
is at the non-rotating model. Thus the evolutionary path of
rapid spin-down extends down to the non-rotating limit. For a massive
remnant with carbon+oxygen envelope, the spin-down path
is much shorter for the Kepler law and there is no spin-down path
for the uniform rotation.

On the other hand, Yoon07 rotation law produces distinct sequences.
We have two branches of evolutionary paths of rapid spin-down.
One is a path connecting the non-rotating limit to the energy maximum.
Another is a sequence starting from the maximum of angular momentum
to the mass-shedding limit. The fate of a remnant at the end of
its rapid spin-down era is thus completely different depending on
which branch it is on. For the former path, the remnant may lose
all the angular momentum (though we have not specified the mechanism
of angular momentum loss). For the latter case, the remnant may shed its 
mass at the equator at the end of spin-down.

We also consider the sequence on which the total angular momentum
is conserved, while the specific entropy decreases. This is meant
to mimic the rapid cooling and slow spin-down evolution.
A branch of the sequence on which total energy decreases
as the entropy decreases is regarded as the evolutionary path.
The uniform rotation and Kelpler sequences are again similar though
the latter sequence extend to larger value of entropy. They are monotonic
in $S_{\rm b}$ thus the sequences are regarded as rapid cooling sequences.
For the Yoon07 rotation law,  the nature of the path strongly depends on 
the mass and the composition of the envelope. For a helium envelope 
case, the massive remnant has an energy minimum on the sequence
(Fig.\ref{fig: M and J const}) while the energy is monotonic for the less
massive case (Fig.\ref{fig: M0.5 and J const}). For the former case,
a remnant born at a point on the right of the energy minimum rapidly
cools down to the minimum. Beyond that the rapid cooling is no
longer valid. For the latter case, the total energy is a decreasing
function of the entropy. Thus the sequence is not regarded as
an evolutionary path due to rapid cooling.

\section*{Acknowledgements}
I thank the anonymous reviewer for his/her careful reading and suggestions that 
help to improve the paper. This work was supported by JSPS Grant-in-Aid for Scientific 
Research(C) 18K03641.

\bibliographystyle{mnras}

\bibliography{pap}
%


%


\bsp	
\label{lastpage}
\end{document}